\documentclass[twocolumn,showpacs]{revtex4}
\usepackage{graphicx,epsf,bm,amsmath,color}
\def\bk{\mbox{\boldmath $k$}}
\def\bq{\mbox{\boldmath $q$}}
\def\br{\mbox{\boldmath $r$}}
\def\bfell{\mbox{\boldmath $\ell$}}
\def\bfsigma{\mbox{\boldmath $\sigma$}}
\def\bftau{\mbox{\boldmath $\tau$}}
\begin{document}
%\begin{CJK*}{SJIS}{}
%\draft
\title{Strength of reduced two-body spin-orbit interaction
from chiral three-nucleon force}
\author{M. Kohno}
\affiliation{Physics Division, Kyushu Dental College,
Kitakyushu 803-8580, Japan}

\begin{abstract}
The contribution of a chiral three-nucleon force to the strength of
an effective spin-orbit coupling is estimated. We first construct
a reduced two-body interaction by folding one-nucleon degrees of
freedom of the three-nucleon force in nuclear matter. The spin-orbit
strength is evaluated by a Scheerbaum factor obtained by the $G$-matrix
calculation in nuclear matter with the two-nucleon interaction plus
the reduced two-nucleon interaction. The problem of the insufficiency
of modern realistic two-nucleon interactions to account for the
empirical spin-orbit strength is resolved. It is also indicated that
the spin-orbit coupling is weaker in the neutron-rich environment.
Because the spin-orbit component from the three-nucleon force is
determined by the low-energy constants fixed in the two-nucleon
sector, there is little uncertainty in the present estimation.
\end{abstract}

\maketitle
%\end{CJK*}

Spin-orbit field in atomic nuclei is essential to reproduce
well-established single-particle shell structure. The empirical
strength of the spin-orbit potential, however, has not been fully
understood on the basis of the realistic nucleon-nucleon force.
The possible role of intermediate isobar $\Delta$-excitation to
the nuclear spin-orbit field was considered in parallel with
the construction of the two-pion-exchange three-nucleon force
(3NF) by Fujita and Miyazawa \cite{FM57}. The problem was reinvestigated
in the early 1980s \cite{OTT80,AB81} to search for the additional
spin-orbit strength. Later, the Illinois group showed \cite{PP93} that
their 3NF makes a substantial contribution to the spin-orbit splitting
in $^{15}$N.

Kaiser and his collaborators investigated, in their several papers
\cite{KFW,NK03,NK04}, the nuclear spin-orbit coupling in the framework
of chiral perturbation theory. The large contributions generated by
iterated one-pion exchange and the 3NF almost
cancel each other \cite{KFW,NK03}, and the short-range spin-orbit
strength in the form of the effective four-nucleon contact-coupling
deduced from realistic nucleon-nucleon interactions accounts
well \cite{NK04} for the empirical one. Because the contact-interaction
in the chiral perturbation, however, is still needed to be regulated
for the application to low-energy nuclear structure calculations,
and their arguments for various contributions seem not to be fully
unified, it is worthwhile to analyze the effective strength of the
spin-orbit coupling by applying the established microscopic theory,
namely the lowest-order Brueckner theory, to the two-nucleon and
three-nucleon interactions in the chiral effective field theory (Ch-EFT).
 
The Thomas form of an average single-particle spin-orbit potential
has been used to describe nucleon spin-orbit coupling:
\begin{equation}
U_{\ell s}^0 \frac{1}{r}\frac{d\rho(r)}{dr}\bfell\cdot \bfsigma,\label{eq:tf}
\end{equation}
where the radial function $\rho(r)$ is a nucleon total density
distribution. The relation of the strength $U_{\ell s}^0$ to a two-body
effective spin-orbit interaction was derived by Scheerbaum \cite{SCH76b}.
By defining the constant $B_S(\bar{q})$ for the triplet odd component
of the effective two-body spin-orbit interaction $v_{\ell s}^{3O}(r)$
\begin{equation}
 B_S(\bar{q})=-\frac{2\pi}{\bar{q}}\int_0^{\infty} dr r^3
 j_1(\bar{q}r)v_{\ell s}^{3O}(r),
\end{equation}
with $j_1$ being a spherical Bessel function,
the single-particle spin-orbit potential for spin-saturated nuclei
may be written as
\begin{equation}
U_{\ell s,\tau}(r)=\frac{1}{2}B_S(\bar{q}) \frac{1}{r}
\frac{d\{\rho(r)+\rho_\tau(r)\}}{dr}\bfell\cdot \bfsigma,
\end{equation}
where $\tau$ specifies either a proton or neutron. We refer to
$B_S(\bar{q})$ as a Scheerbaum factor, which is different from the
original constant in Ref. \cite{SCH76b} by a factor of $-\frac{2\pi}{3}$.
Scheerbaum prescribed $\bar{q}\approx 0.7$ fm$^{-1}$ on the basis of the
wavelength of the density distribution. We employ this prescription.
If we assume a naive relation $\rho_p(r)=\rho_n(r)=\frac{1}{2}\rho(r)$,
we recover the Thomas form, Eq. (\ref{eq:tf}),
with $U_{\ell s}^0=\frac{3}{4}B_S(\bar{q})$. It has also been customary
to use a $\delta$-type two-body spin-orbit interaction
\begin{equation}
iW(\bfsigma_1+\bfsigma_2)\cdot (\nabla_r\times \delta(\br)\nabla_r)
\end{equation}
in nuclear Hartree-Fock calculations using $\delta$-type Skyrme interactions
\cite{SKY75,SKY80} and even with finite range effective forces,
e.g., the Gogny force \cite{GP77}. This two-body force provides
a single-particle spin-orbit potential:
\begin{equation}
 \frac{1}{2}W \frac{1}{r}\frac{d\{\rho(r)+\rho_\tau(r)\}}{dr}\bfell\cdot\bfsigma.
\end{equation}
Therefore, the strength $W$ may be identified as the Scheerbaum factor
$B_S(\bar{q})$. The empirical value of $W$ is around $120$ MeV$\cdot$fm$^5$
in various nuclear Hartree-Fock calculations. As will be shown below,
the modern nucleon-nucleon interactions underestimate the spin-orbit
strength by about 25 \%.

Applying Scheerbaum's formulation to the momentum-space $G$-matrix
calculation in nuclear matter with the Fermi momentum $k_F$,
we obtain the corresponding spin-orbit strength as follows \cite{FK00}:
\begin{eqnarray}
 B_S(\bar{q})=\frac{1}{k_F^3}\sum_{JT}(2J+1)(2T+1)\int_0^{q_{max}} dq \nonumber\\
 \times W(\bar{q},q)\{ (J+2)G_{1J+1,1J+1}^{JT}(q)+G_{1J,1J}^{JT}(q) \nonumber \\
 -(J-1)G_{1J-1,1J-1}^{JT}(q) \}.
\end{eqnarray}
Here, $q_{max}=\frac{1}{2}(k_F+\bar{q})$ and the weight factor $W(\bar{q},q)$ is
\begin{equation}
 W(\bar{q},q)=\left\{ \begin{array}{l}
 \theta (k_F-\bar{q})\;\;\mbox{for} \;\; 0\leq q\leq \frac{|k_F-\bar{q}|}{2} \\
 \frac{k_F^2-(\bar{q}-2q)^2}{8\bar{q}q}\;\;\mbox{for}\;\;
 \frac{|k_F-\bar{q}|}{2} \leq q\leq \frac{k_F+\bar{q}}{2},
 \end{array} \right.
\end{equation}
where $\theta (k_F-\bar{q})$ is a step function.
In Eq. (6), $G_{1\ell',1\ell}^{JT}$ is the abbreviation of the momentum-space
diagonal $G$-matrix element in the spin-triplet channel with the total
isospin $T$, total spin $J$, and orbital momenta $\ell'$ and $\ell$. 

Calculating $B_S(\bar{q})$ in the lowest-order Brueckner theory with the
continuous prescription for intermediate spectra, as presented below explicitly
in Table 1, modern two-body nucleon-nucleon potentials are found to give
smaller values of around 90 Mev$\cdot$fm$^5$ compared with the empirical one. 
As has been well known that
LOBT calculations in symmetric nuclear matter with realistic two-nucleon force
do not reproduce correct saturation property. However, in most case, calculated
energies at the empirical saturation point $k_F=1.35$ fm$^{-1}$ are close to the
empirical energy of about $-16$ MeV. This suggests that $G$ matrices
provide basic information on the effective nucleon-nucleon interaction
in the nuclear medium, by incorporating important short-range correlations,
Pauli effects and dispersion effects.

Now we consider the contribution of the 3NF. In this article, we estimate
it in a two-step procedure. First, the 3NF $v_{123}$ defined in momentum
space is reduced to an effective two-nucleon interaction $v_{12(3)}$ by
folding one-nucleon degrees of freedom:
\begin{widetext}
\begin{equation}
 \langle \bk_1' \sigma_1' \tau_1', \bk_2'\sigma_2'\tau_2'|v_{12(3)}
 |\bk_1 \sigma_1 \tau_1, \bk_2\sigma_2\tau_2\rangle_A
 =\frac{1}{3}\sum_{\bk_3\sigma_3\tau_3} \langle \bk_1'\sigma_1'\tau_1',
 \bk_2'\sigma_2'\tau_2', \bk_3\sigma_3\tau_3|v_{123}|\bk_1\sigma_1\tau_1,
 \bk_2\sigma_2\tau_2, \bk_3\sigma_3\tau_3\rangle_A.\label{eq:efv}
\end{equation}
\end{widetext}
Here, we have to assume that remaining two nucleons are in the
center-of-mass frame, namely $\bk_1'+\bk_2'=\bk_1+\bk_2=0$.
The density-dependent effective two-nucleon interaction as the
effect of the 3NF has been commonly introduced in the literature
\cite{KAT74,FP81,HKW10}. Note that the suffix $A$ means an
antisymmetrized matrix element; namely $|ab\rangle_A\equiv |ab-ba\rangle$
and $|abc\rangle_A\equiv |abc-acb+bca-bac+cab-cba\rangle$,
 and the factor $\frac{1}{3}$ in Eq. (\ref{eq:efv}) is an additional statistical
one. This statistical factor has been often slipped
in the literature. The recent derivation of the effective two-body
interaction from the Ch-EFT 3NF by Holt, Kaiser and Weise \cite{HKW10}
also seems not to be an exception. If an adjustable strength is introduced,
the statistical factor may be hidden in the fitting procedure.
In our case of using the Ch-EFT 3NF, the low-energy constants except
for $c_D$ and $c_E$ are fixed. Although there may be a room to adjust
$c_D$ and $c_E$, the contributions to the energy from these terms are
rather small, if they are in a reasonable range. In addition, $c_D$ and
$c_E$ do not contribute to the reduce two-nucleon spin-orbit interaction.
By comparing the nuclear matter energy directly calculated from $v_{123}$
and that by the reduced $v_{12(3)}$, the error
due to this approximation can be checked to be less than 10 \%, if we
calculate Born energy without including a form factor.

To explain the procedure of obtaining $v_{12(3)}$ more explicitly, we write
the reduced spin-orbit component originating from the $c_1$ term of the
Ch-EFT 3NF:
\begin{equation}
 -\frac{c_1 g_A^2 m_\pi^2}{f_\pi^4}\sum_{1\leq i<j\leq 3}
 \frac{(\bfsigma_i\cdot \bq_i)(\bfsigma_j\cdot\bq_j)}
 {(\bq_i^2+m_\pi^2)(\bq_j^2+m_\pi^2)}(\bftau_i\cdot\bftau_j),
\end{equation}
where $g_A=1.29$, $f_\pi =92.4$ MeV, $m_\pi$ is a pion mass, and $\bq_i$
is a momentum transfer of the $i$-th nucleon. The momentum
transfer of the third nucleon $k$ is dictated by the relation
$\bq_k=-\bq_i-\bq_j$. The folding of the 3NF by
one nucleon is carried out without incorporating a three-body form
factor. A form factor is later introduced on the two-body level. The
folding in symmetric nuclear matter with the Fermi momentum $k_F$ gives,
besides the central and tensor components, the following spin-orbit term:
\begin{eqnarray}
 & & \frac{c_1 g_A^2 m_\pi^2}{f_\pi^4}\frac{1}{(2\pi)^3}\iiint_{|\bk_3|\le k_F}
    d\bk_3 \nonumber\\
 & & \times \frac{i(\bfsigma_1+\bfsigma_2)\cdot
 (-\bk_1'\times \bk_1+(\bk_1'-\bk_1)\times\bk_3)}
 {((\bk_1'-\bk_3)^2+m_\pi^2)((\bk_1-\bk_3)^2+m_\pi^2)}.\label{eq:nmc1}
\end{eqnarray}
When carrying out the folding in pure neutron matter, the restriction of
the isotopic spin brings about an additional factor of $\frac{1}{3}$.

The partial-wave decomposition of the above spin-orbit term becomes
\begin{eqnarray}
 & & -\delta_{S 1} \frac{c_1 g_A^2 m_\pi^2}{f_\pi^4}
 \frac{\ell(\ell+1)+2-J(J+1)}{2\ell+1} \nonumber \\
 & &  \left\{ Q_{W,0}^{\ell-1}(k_1',k_1)-Q_{W,0}^{\ell+1}(k_1',k_1)
 -W_{\ell s,0}^\ell (k_1',k_1)\right\}
\end{eqnarray}
for the orbital and total angular momenta $\ell$ and $J$.
The functions $Q_{W,0}^\ell$ and $W_{\ell s,0}^\ell$ are defined by
\begin{eqnarray}
 Q_{W,0}^\ell (k_1',k_1)\!&\!\equiv\!& \!\frac{2\pi}{(2\pi)^3}\frac{1}{2}
 \int_0^{k_F}\!dk_3 Q_\ell (x')Q_\ell (x),\\
 W_{\ell s,0}^\ell (k_1',k_1)\!&\!\equiv\!&\!\frac{2\pi}{(2\pi)^3}
 \frac{1}{2k_1'k_1}\int_0^{k_F}\!dk_3k_3 \nonumber \\
 & & \times\left\{ k_1'Q_\ell (x)(Q_{\ell-1}(x')-Q_{\ell+1}(x'))\right.\nonumber\\
 & &\left. \!\!+ k_1Q_\ell (x')(Q_{\ell-1}(x)-Q_{\ell+1}(x))\right\},
\end{eqnarray}
where $Q_\ell(x)$ is a Legendre function of the second kind,
and $x'\equiv \frac{k_3^2+k_1'^2+m_\pi^2}{2k_1'k_3}$
and $x\equiv \frac{k_3^2+k_1^2+m_\pi^2}{2k_1k_3}$, respectively.

The spin-orbit component arises also from the $c_3$ term of the Ch-EFT 3NF.
This case, in addition to the replacement of the coupling constant, an
additional factor $(\bk_1'-\bk_3)\cdot (\bk_3-\bk_1)$ appears in the
denominator in Eq. (\ref{eq:nmc1}). The partial-wave decomposition reads
\begin{widetext}
 \begin{eqnarray}
  \delta_{S 1}\frac{c_3g_A^2}{2f_\pi^4} \frac{\ell(\ell+1)+2-J(J+1)}{2\ell+1}
  \left[ (m_\pi^2+\frac{1}{2}(k_1'^2+k_1^2))\{ Q_{W,0}^{\ell-1}(k_1',k_1)-
  Q_{W,0}^{\ell+1}(k_1',k_1)-W_{\ell s,0}^{\ell}(k_1',k_1)\}\right.\nonumber\\
  +3k_1'k_1 \left\{ Q_{W,0}^\ell (k_1',k_1)-(\ell-1)Q_{W,0}^{\ell-2}(k_1',k_1)
 +(\ell +2)Q_{W,0}^{\ell+2}(k_1',k_1) +\frac{\ell-1}{2\ell-1}
  W_{\ell s,0}^{\ell-1}(k_1',k_1)
 +\frac{\ell+2}{2\ell+3} W_{\ell s,0}^{\ell+1}(k_1',k_1)\right\}\nonumber\\
  \left. -\delta_{\ell 1} \frac{k_1'k_1}{2}
  (F_0(k_1')+F_0(k_1)-F_1(k_1')-F_1(k_1)) \right],
 \end{eqnarray}
\end{widetext}
where the new functions $F_0(k)$ and $F_1(k)$ are defined by
\begin{eqnarray}
 F_0(k) \equiv \frac{1}{(2\pi)^3}\iiint_{|\bk_3|\leq k_F}d\bk_3
 \frac{1}{(\bk-\bk_3)^2+m_\pi^2},\\
 F_1(k) \equiv \frac{1}{(2\pi)^3}\frac{1}{k^2}\iiint_{|\bk_3|\leq k_F}d\bk_3
 \frac{\bk\cdot \bk_3}{(\bk-\bk_3)^2+m_\pi^2}.
\end{eqnarray}

Adding the reduced two-nucleon interaction to the Ch-EFT two-nucleon
interaction, we repeat the LOBT $G$-matrix calculation. Although explicit
expressions are not shown in this Letter except for the spin-orbit part, we
include all central, tensor and spin-orbit components of the reduced
interaction $v_{12(3)}$. The form factor in a functional form of
$f(k_1',k_1)=\exp\{-[(k_1'/\Lambda)^4+(k_1/\Lambda)^4]\}$ is introduced for
$v_{12(3)}$ with the cut-off mass $\Lambda=$550 MeV. We use the low-energy
constants fixed for the J\"{u}lich Ch-EFT potential by Hebeler {\it et al.}
\cite{HEB}; $c_D=-4.381$, and $c_E=-1.126$. Other constants are $c_1=-0.81$
GeV$^{-1}$, $c_3=-3.4$ GeV$^{-1}$, and $c_4=3.4$ GeV$^{-1}$. Because the
reduction of the 3NF to the two-nucleon force was carried out in nuclear
matter, $v_{12(3)}$ may not be directly applied to very light nuclei,
such as $^3$H and $^4$He.

First, we comment on calculated saturation curves, which are given in Fig. 1.
Without the contribution of the 3NF, the saturation curve attains its minimum
at larger $k_F$ as a function of the Fermi momentum $k_F$ than the empirical
saturation momentum, as has been known. Nucleon-nucleon interactions,
AV18 \cite{AV18}, NSC97 \cite{NSC}, and J{\"{u}}lich N$^3$LO with the
cutoff mass of 550 MeV \cite{EHM09} give similar saturation curves, and the
CD-Bonn potential \cite{CDB} predicts somewhat deeper binding. For the
reference of what saturation curve is preferable for nuclear mean filed
calculations, we also show the result with the Gogny D1S interaction \cite{GP77}.

The thin dotted curve shows the result in which the plane wave expectation
value of the 3NF $v_{123}$ is added to the result of the two-nucleon N$^3$LO.
The thick dotted curve alongside the thin dotted curve is the result with
the plane wave expectation value of the reduced two-nucleon interaction
$v_{12(3)}$. The difference between the thin and thick curves is due to
the difference of the form-factors and the necessary approximation
$\bk_1'+\bk_2'=\bk_1+\bk_2=0$ in Eq. (8).

The solid curve is the result of the $G$-matrix calculation with including
the reduced two-nucleon interaction, $v_{12(3)}$. Although the energy is
seen to be underestimated by a few MeV, the saturation property is largely
improved by the repulsive contribution from the three-nucleon force.
It is not necessary at present to expect a perfect agreement with the
empirical properties in the LOBT calculation in nuclear matter.

\begin{figure}
\includegraphics[width=75mm]{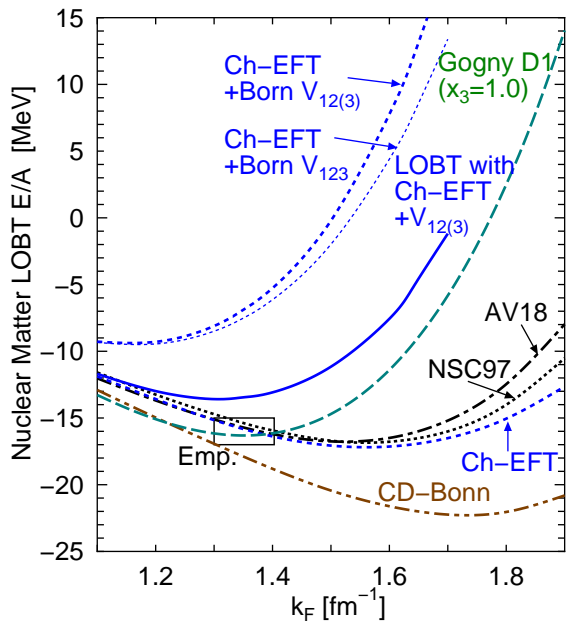}
\caption{Saturation curves in symmetric nuclear matter.
}
\end{figure}
 
\begin{table}[t]
\begin{tabular}{crrrrr}\hline \hline
 $k_F=1.35$ fm$^{-1}$ & AV18 & NSC97 & CD-B & N$^3$LO & N$^3$LO+3NF \\ \hline\hline
 $B_S(T=0)$ & 2.09 & 1.9 & 3.1 & 2.5 & 7.0 \\
 $B_S(T=1)$ & 86.4 & 86.7 & 90.2 & 84.6 & 116.2 \\ \hline
 $B_S(\bar{q})$ & 88.4 & 88.6 & 93.3 & 87.1 & 123.2 \\ \hline \hline
$k_F=1.07$ fm$^{-1}$ & AV18 & NSC97 & CD-B  & N$^3$LO & N$^3$LO+3NF \\ \hline
 $B_S(T=0)$ & 1.4  & 1.3 & 2.3 & 1.6 & 4.1 \\
 $B_S(T=1)$ & 88.1  &  88.7 & 92.2 &  86.5 & 106.7 \\ \hline
 $B_S(\bar{q})$ & 89.5 & 90.0 & 94.5 & 88.1 & 110.8 \\ \hline \hline
\end{tabular}
\caption{$B_S(\bar{q})$ in the unit of MeV$\cdot$fm$^5$
given by Eq. (6) with $\bar{q}=0.7$ fm$^{-1}$ for modern nucleon-nucleon
interaction: AV18 \cite{AV18}, NSC97 \cite{NSC},  CD-Bonn \cite{CDB},
and J{\"{u}}lich N$^3$LO \cite{EHM09}.
The last entry is the result with including the reduced two-body interaction
from the Ch-EFT 3NF.}
\end{table}

Now we examine the spin-orbit strength. We tabulate values
for $B_S(\bar{q})$ of Eq. (6) at $\bar{q}=0.7$ fm$^{-1}$ calculated in
the LOBT with modern nucleon-nucleon interactions: AV18 \cite{AV18},
NSC97 \cite{NSC}, CD-Bonn \cite{CDB}, and J{\"{u}}lich N$^3$LO \cite{EHM09}.
The Scheerbaum factors obtained by realistic two-nucleon forces are seen
to be similar but insufficient to explain the strength needed in nuclear
mean field calculations. Namely only about three-fourths of the empirically
needed strength is accounted for. The two-body part of the Ch-EFT, N$^3$LO,
shows little difference with other realistic two-nucleon force.
It is also noticed that values at $k_F=1.07$ fm$^{-1}$, namely at the half
of the normal density, change little from those at the normal density with
$k_F=1.35$ fm$^{-1}$. It turns out, as the last column
of Table I shows, that the addition of the reduced two-body interaction from
the Ch-EFT 3NF bring about a good effect to fill the gap, though the 3NF
contribution is smaller at $k_F=1.07$ fm$^{-1}$. This is in accord with the
important role of the 3NF to the spin-orbit splitting demonstrated in quantum
Monte Carlo calculations of low-energy neutron-alpha scattering \cite{NOL}.
Although there are ambiguities from the form factor and uncertainties
inherent in the folding procedure without taking into account
nucleon-nucleon correlations, no additional adjustable parameter exists,
because low-energy constants $c_1$ and $c_3$ which contribute solely to
the spin-orbit strength are determined on the two-nucleon sector.

As noted after Eq. (\ref{eq:nmc1}), the reduced two-body spin-orbit term
in neutron matter is one-third of that in symmetric nuclear matter.
Actual $G$-matrix calculations using the Ch-EFT N$^3$LO plus $v_{12(3)}$
in pure neutron matter with $k_F^n=1.35$ fm$^{-1}$ tell that $B_S(\bar{q})$
values at $\bar{q}=0.7$ fm$^{-1}$ are 84.7 and 93.5 MeV$\cdot$fm$^5$
without and with the reduced two-nucleon interaction $v_{12(3)}$,
respectively. If $k_F^n=1.07$ fm$^{-1}$ is assumed, the corresponding
values are 87.0 and 94.6 MeV$\cdot$fm$^5$, respectively.
Again, the $k_F^n$-dependence is weak.
While the spin-orbit strength from the two-nucleon force is
scarcely different from that in symmetric nuclear matter, the
additional contribution from the three-nucleon force
is in fact almost one-third of that in symmetric nuclear matter.
Thus, the spin-orbit strength is expected to be smaller in the neutron-rich
environment. This seems to be consistent with the trend
observed in the shell structure near the neutron drip line \cite{SCH04}
that a decreasing spin-orbit interaction
is preferable with increasing neutron excess.

In summary, we have estimated quantitatively the contribution  of the
three-nucleon force of the chiral effective field theory to the single-particle
spin-orbit strength, using the formulation by Scheerbaum \cite{SCH76b}.
We first introduced the reduced two-body interaction by folding one-nucleon
degrees of freedom of the 3NF in nuclear matter. Making partial-wave expansion
of the resulting two-body interaction and adding it to the genuine two-nucleon
interaction with including the necessary statistical factor of $\frac{1}{3}$,
we carried out LOBT $G$-matrix calculations in infinite matter and
evaluated the Scheerbaum factor corresponding to the spin-orbit strength.
Because the spin-orbit field in the atomic nuclei is fundamentally important
as the nuclear magic numbers exhibit, it is important to learn
that the inclusion of the 3NF in the chiral effective field theory can
account for the spin-orbit strength empirically required for nuclear mean
filed calculations. Because the relevant low-energy constants $c_1$ and
$c_3$ are determined in the two-nucleon interaction sector, there should
be little uncertainty for the additional spin-orbit strength except for
the treatment of the two-body form factor. We have also noted that the
additional spin-orbit strength from the 3NF should be weaker in
neutron-excess nuclei.

\acknowledgments
This work is supported by Grant-in-Aid for Scientific Research (C) from
the Japan Society for the Promotion of Science (Grant No. 22540288).
The author thanks H. Kamada for valuable comments concerning the Ch-EFT
interaction. He is also grateful to M. Yahiro for his interest in this work.

\end{document}